\input{psfig.sty}
\documentclass[11pt,a4paper]{article}
\usepackage{jcappub}
\usepackage{subfigure}
\newcommand{\goodgap}{\hspace{\subfigtopskip} \hspace{\subfigbottomskip}}

\title{The flat density profiles of massive, and relaxed galaxy clusters}
\author[a,b]{A. Del Popolo}
\affiliation[a]{Dipartimento di Fisica e Astronomia, University Of Catania, \\
Viale Andrea Doria 6, 95125 Catania, Italy \\}
\affiliation[b]{International Institute of Physics, Universidade Federal do Rio Grande do Norte,\\
59012-970 Natal, Brazil}

\emailAdd{adelpopolo@oact.inaf.it}
\abstract{The present paper is an extension and continuation of Del Popolo (2012a) which studied the role of baryon physics on clusters of galaxies formation.
In the present paper, we studied by means of the SIM introduced in Del Popolo (2009), the total and DM density profiles, and the correlations among different quantities, observed by Newman et al. (2012a,b), in seven massive and relaxed clusters, namely MS2137, A963, A383, A611, A2537, A2667, A2390.  
As already found in Del Popolo 2012a, the density profiles depend on baryonic fraction, angular momentum, and the angular momentum transferred from baryons to DM through dynamical friction.
Similarly to Newman et al. (2012a,b), the total density profile, in the radius range 0.003 - 0.03$r_{200}$, has a mean total density profile in agreement with dissipationless simulations.
The slope of the DM profiles of all clusters is flatter than -1. The slope, $\alpha$, has a maximum value (including errors) of $\alpha=-0.88$ in the case of A2390, and minimum value $\alpha=-0.14$ for A2537. The baryonic component dominates the mass distribution at radii $< 5-10$ kpc, while the outer distribution is dark matter dominated. 
We found an anti-correlation among the slope $\alpha$, the effective radius, $R_e$, and the BCG mass, and a correlation among the core radius $r_{core}$, and $R_e$. Moreover, the mass in 100 kpc (mainly dark matter) is correlated with the mass inside 5 kpc (mainly baryons). 

The behavior of the total mass density profile, the DM density profile, and the quoted correlations can be understood in a double phase scenario. In the first dissipative phase the proto-BCG forms, and in the second dissipationless phase, dynamical friction between baryonic clumps (collapsing to the center) and the DM halo flattens the inner slope of the density profile. 

In simple terms, the large scatter in the inner slope from cluster to cluster, and the anti-correlation among the slope, $\alpha$ and $R_e$ is due to the fact that in order to have a total mass density profile which is NFW-like, clusters having more massive BCGs at their centers must contain less DM in their center. Consequently the inner profile has a flatter slope. 

}

\begin{document}
\maketitle

\section{Introduction}

A fundamental prediction and test of the CDM paradigm is the structure and abundance of dark matter (DM) haloes (Del Popolo \& Gambera 2000; Del Popolo 2007). The density profiles predicted in dissipationless N-body simulations (Navarro, Frenk, \& White (1996, 1997) (hereafter NFW), Power et al. (2003), Navarro et al. (2004); Diemand et al. 2005) have a central density cusp, characterized by $\rho_{DM} \simeq r^{\alpha} \simeq r^{-1}$. The cusp is shallower at smaller scales (Navarro et al. 2010) with $\alpha \simeq 0.8$ at 120 pc (Stadel et al. 2009)\footnote{For precision's sake, we remind that Moore et al (1998), and Fukushige \& Makino (2001) found a steeper cusp, $\rho_{DM} \simeq r^{-1.5}$, and Ricotti 2003; Ricotti \& Wilkinson,
2004; Ricotti Pontzen, \& Viel 2007; Del Popolo 2010 (DP10), Del Popolo 2011 (DP11) found different values of the inner density profile slope according to the objects considered (galaxies, clusters).}. Observations of dwarf galaxies, low-surface-brightness galaxies (LSBs), objects which are DM dominated, are at odds with simulations, finding core-like density profiles (e.g. Flores \& Primack 1994; Moore 1994; de Blok\& Bosma 2002; de Blok, Bosma \& McGaugh 2003;  Gentile et al. 2004, 2006; Span\'o et al. 2008; Kuzio de Naray et al. 2008, 2009; Oh et al. 2010; Del Popolo \& Hiotelis 2014)\footnote{The $\Lambda$CDM model suffers from other small scale problems (Del Popolo \& Gambera 1997; Del Popolo et al. 2014), and other problems like the cosmological constant problem (Weinberg 1989; Astashenok, \& Del Popolo 2012), and the cosmic coincidence problem".
Moreover, the dark energy is assumed to be related to the cosmological constant, while several other possibilities for the dark energy have been proposed (Del Popolo, Pace, Lima 2013a,b; Del Popolo et al. 2013)}. 

While in spiral galaxies the determination of the density profile is simplified by the presence of HI, and the determination of the galaxy rotation curve, in different structures (e.g., dsPh galaxies, ellipticals) the determination of the DM mass profile is challenging, because of 
the small dynamic range of observations,
the degeneracies related to the mass probes used in the profile determination (see Del Popolo 2002) (e.g., velocity anisotropy), and the difficulty in disentangling baryons and DM (see Del Popolo 2013, section ``Dark matter distribution", and Del Popolo 2014). The quoted problems can be overcome in clusters of galaxies, since clusters have properties that 
can be understood and interpreted in a simpler fashion than galaxy rotation curves.
Several observational probes (e.g., X-ray emission from intra-cluster plasma, gravitational lensing) furnish an accurate measure of mass (e.g. Allen et al. 2011; Kneib \& Natarajan 2011). 

In the last years, studies of clusters of galaxies, showed that the total inner density profile is well described by dissipationless N-body simulations at radii $\geq 5-10$ kpc, while the DM profiles are flatter than those obtained in the simulations (Sand et al 2002, 2004, 2008 (Sa02; Sa04, Sa08); Newman et al. 2009, 2011, 2013a,b (N09, N11, N13a,b)), within a radius of $\simeq 30$ kpc, typical of the BCG radius. Moreover, the DM profile is characterized by a variation of the slope, $\alpha=-d \log{\rho_{DM}}/ d r$, from cluster to cluster, and the variation correlates with the Brightest Cluster Galaxy (BCG) properties. 

A similar scatter in $\alpha$ from galaxy to galaxy was also observed by Simon et al. (2003, 2005) in the dwarf galaxies sample constituted by NGC 2976, NGC 4605, NGC 5949, NGC 5963, and NGC 6689, having $\alpha=(0.01; 0.78; 0.88; 1.20; 0.79)$, respectively, mean slope $\alpha \simeq 0.73$, and a dispersion of 0.44.


The discrepancy among the inner slope predicted by dissipationless simulations and observations is interpreted in terms of the fact that dissipationless simulations are not taking into account baryons that are of fundamental importance in the inner parts of galaxies and clusters (see N13b). 
The quoted discrepancy is reduced or eliminated, both in galaxies and clusters of galaxies, when baryons are taken into account by means of SPH simulations or by semi-analytic models (El-Zant et al. 2001, 2004; Nipoti et al. 2004; Jardel \& Sellwood 2009; Del Popolo \& Kroupa (2009); Del Popolo 2009 (DP09), DP10, DP11, DP12a,b; Cardone \& del Popolo 2012; Del Popolo, Cardone, Belvedere 2013; Governato et al. 2010, 2012; Cole et al. 2011; Cardone et al. 2011; Martizzi et al. 2012).

In DP12a, we used the model introduced in DP09 to study how baryonic physics influence the shape of the clusters density profiles, showing that the baryon presence in the inner 10 kpc of the structure modifies the inner profile, and finding correlation among the inner baryon content (mass of the BCG) and the slope of the density profile. Moreover, we studied the density/mass profiles of some clusters (A611, A383, MACS J1423.8+2404, RXJ1133) previously studied by other authors (Sand et al. 2004; Newman et al. 2009, 2011; Morandi, Pedersen \& Limousin 2010), finding a good agreement with the observed profiles.  

Meanwhile, N13b used a larger sample\footnote{The seven clusters constituting the sample are: MS2137, A963, A383, A611, A2537, A2667, A2390.}
than those used in previous studies, 
to obtain a 
joint measurement of the stellar mass scale,
and improving the analysis performed in the previous papers. 
They obtained the total and DM density profile for the quoted clusters, finding that the average total density is compatible with a NFW profile, and the DM profile of all clusters is shallower than what dissipationless simulations indicate. Moreover, they found correlations among the slope of the inner part of the DM density profile and some BCG structure parameters.

N13a,b presented an intuitive interpretation of the results and correlation found in the observations, proposing a ``physical picture" based on previous theoretical results (e.g., El-Zant et al. 2001, 2004; Nipoti et al. 2004; DP09; Laporte et al. 2012; DP12a). The ``physical picture" proposed by N13a,b, is based on a first dissipative phase in which the BCG forms, and a second dissipationless phase in which dynamical friction between baryons clumps (collapsing to the center of the proto-structure) and the DM halo reduces the central DM density.  

Then, it would be very interesting to see if the quoted correlations can be re-obtained using a theoretical model, to understand what causes them, and eventually to confirm if the ``physical picture" proposed by N13a,b is consistent.  

To this aim, we use the new observations and new data published in N13a,b, in the model of DP09, similarly to what was done in DP12a.
DP09 accounts, among other effects, for the adiabatic contraction giving rise to a steepening of the inner density slope, and for the inner cusp ``heating" produced by the exchange of angular momentum from baryons to the DM through dynamical friction. We already studied in DP12a, how clusters density profiles are shaped by the interplay among baryons and DM.
In that paper, we showed that halos containing only DM, like in dissipationless simulations, have Einasto's profiles, while if baryons are taken into account, the profiles flatten proportionally to the baryons content (especially the central baryon content). 
We applied the theoretical results of DP12a to reproduce the mass and density profiles of some well studied clusters. Moreover, we found correlations among the inner slope of clusters, their total baryons content, and the mass content of the inner 10 kpcs.

In the present paper, similarly to what done in DP12a, we will study the density profiles and correlations found from the new data by N13a,b, and compare them with N13a,b observations. 

The plan of the paper is the following. In section 2, we give a summary of the model. In section 3, we discuss the results, and section 4 is devoted to conclusions.

\section{Summary of the model}

The model used is described in DP09, and DP12a,b. We refer readers to those papers for details, while here we give a summary of the model. 

To start with, DP09 is an improved spherical infall model (SIM), that differently from previous SIMs (Gunn \& Gott 1972;  Hoffman \& Shaham 1985; Ryden \& Gunn 1987; Ascasibar, Yepes \& G\"ottleber 2004; Williams, Babul \& Dalcanton 2004), includes simultaneously the effect of DM adiabatic contraction, those of random and ordered angular momentum, and angular transfer among baryons and DM through
dynamical friction. The quoted model differs from previous SIMs for the fact that it simultaneously takes into account the effects that previous SIMs 
accounted one at a time. Namely, random angular momentum (e.g., Williams, Babul \& Dalcanton 2004), dynamical friction of stellar/DM clumps against the background halo (e.g., El-Zant et al. 2001; Romano-Diaz et al. 2008), and adiabatic contraction (e.g., Blumenthal et al. 1986; Gnedin et al. 2004; Klypin,  Zhao, and Somerville 2002; Gustafsson et al. 2006). 

Following Gunn \& Gott (1972), a protostructure is considered as formed by concentric shells, expanding with the Hubble flow. Starting from an initial comoving radius $x_i$, each shell expands to a maximum radius, $x_m$, usually termed turn-around radius,  $r_{ta}$, and then it collapse giving rise to a  
``virialized" structure, when non-linear processes in the collapse phase converts kinetic energy is converted into random motions (Hiotelis \& Del Popolo 2006, 2013). 

The final density is given by (Gunn 1977, Fillmore \& Goldreich 1984)
\begin{equation}
\rho(x)=\frac{\rho_{ta}(x_m)}{f(x_i)^3} \left[1+\frac{d \ln f(x_i)}{d \ln x_m(xi)} \right]^{-1}
\label{eq:dturnnn}
\end{equation}
where the term $f(x_i)=x/x_m(x_i)$ is the so called collapse factor (see Eq. A18, DP09).

In the original SIM of Gunn \& Gott (1972), the collapse was radial, and did not take into consideration the angular momentum, which originates from tidal interaction of the proto-structure with the neighbors (Hoyle 1953; Peebles 1969; White 1984; Ryden 1988; Eisenstein \& Loeb 1995; Catelan \& Theuns 1996; Schaefer 2009). This ``ordered" angular momentum is obtained integrating the torque over time on each mass shell (e.g., Ryden 1988, Eq. 35).
 
It is usual to express the total angular momentum in terms of a dimensionless quantity, the spin parameter, $\lambda$,
\begin{equation}
\lambda=\frac{L |E|^{1/2}}{GM^{5/2}}=\frac{\omega}{\omega_{sup}}=\frac{L}{2G^{1/2}M^{3/2} R^{1/2}}, 
\end{equation}
(Peebles 1969, Padmannabhan 1993), where $E$ is the halo's binding energy, $L$ the angular momentum, $\omega$ the angular velocity of the system, and $\omega_{sup}$ the angular velocity providing the rotational support.
In a system, like ours, constituted of baryons and DM, the previous equation can be written as
\begin{equation}
\lambda_{gas(DM)}=\frac{L_{gas(DM)}}{M_{gas(DM)} [2G(M_{gas}+M_{DM})r_{vir}^{1/2}]}, 
\end{equation}
where $M_{gas(DM)}$ is the gas(DM) mass contained in the virial radius $r_{vir}$, and $L_{gas(DM)}$ is the angular momentum of gas(DM). Following G\"ottleber \& Yepes (2007), $\lambda_{gas}/\lambda_{DM}=1.23$ for haloes with $M_{vir}> 5 \times 10^{14} h^{-1}M_{\odot}$ (see their figure 5)\footnote{The $\lambda$ parameter is log-normally distributed with $\lambda=0.0351 
\pm 0.0016$, $\sigma_{\lambda}=0.6470 \pm 0.0067$, in the case of DM, and $\lambda=0.0462 \pm 0.0012$, $\sigma_{\lambda}=0.6086 \pm0.0030$ for the gas distribution, and $\lambda_{\max}=0.0231, 0.0319$ for DM and gas distributions (see Bett et al. 2007; Sharma \& Steinmetz 2005).}. 

A ``random" angular momentum, $j$, is also present in haloes, and is generated by random velocities (Ryden \& Gunn 1987). 
It can be taken into account assigning an angular momentum at turn-around
(e.g., Nusser 2001; Hiotelis 2002; Ascasibar, Yepes \& G\"ottleber 2004), 
\begin{equation}
j=
\propto \sqrt{GM x_m}  
\end{equation}
which can also be expressed in terms of the eccentricity ratio $e_0=\left( \frac{r_{min}}{r_{max}} \right)_0$, $r_{max}$, and $r_{min}$ being the apocentric and pericentric radii, respectively. N-body simulations show that $< \frac{r_{min}}{r_{max}}> \simeq 0.2$ in virialized halos (Avila-Reese et al. 1998). Since moving to the turn-around radius, $r_{ta}$, particles orbits are more radial, one needs to use a correction as shown by Ascasibar, Yepes \& G\"ottleber (2004)  
\begin{equation}
e(r_{max}) \simeq 0.8 (r_{max}/r_{ta})^{0.1}
\end{equation}
for $r_{max}< 0.1 r_{ta}$. Then angular momentum can be taken into account using the previous approach: 
Avila-Reese et al. (1998) approach with the Ascasibar, Yepes \& G\"ottleber (2004) correction. 

The deceleration term connected to dynamical friction was introduced in the equation of motion (Eq. A14 in DP09). The dynamical friction coefficient was obtained similarly to Antonuccio-Delogu \& Colafrancesco (1994) (see also Appendix D of DP09).   

The adiabatic contraction (AC) of DM produced by the baryons collapse was taken into account as follows. 
Our protostructure is made of baryons and DM,
baryonic fraction $F_b=M_b/M_{500} <<1$, and DM fraction 
$1-F_b$\footnote{$M_{500}$ is the mass enclosed in a radius $R_{500}$ within which the density is 500 $\rho_c$, being $\rho_c$ is the critical density. 
The total baryonic mass, $M_b$, is given by the sum of the gas mass, $M_{gas}$, and the mass in stars, $M_{\ast}$.}. 
Baryons cools and collapse towards the structure center giving rise to a distribution $M_b( r)$.  
DM is compressed, and particles located initially at $r_i$ move to a new position 
\begin{equation}
r \left [ M_b( r) +M_{dm} ( r) \right] = r_i M_i (r_i)
\label{eq:ad1}
\end{equation}
(Blumenthal et al. 1986), being $M_i (r_i)$ the total mass at initial time, and $M_{dm}$ the final distribution of DM. 
One then assumes that baryons and DM have the same initial distribution (Mo  et al. 1998; Cardone \& Sereno 2005; Treu \& Koopmans 2002; Keeton 2001), and that the final baryon distribution is a Hernquist configuration (Rix et al. 1997; Keeton 2001; Treu \& Koopmans 2002).
If particles orbits do not cross, we have 
\begin{equation}
M_ {dm} ( r)=(1-F_b) M_i (r_i)
\label{eq:ad2}
\end{equation}
Once $M_i (r_i)$ and $M_b (r)$ are given, Eqs. (\ref{eq:ad1}), (\ref{eq:ad2}) can be solved to find the final halo distribution. 
The previous model can be improved assuming that
\begin{equation}
  M(\bar{r})r= {\rm const}.
  \label{eq:modified}
\end{equation}
(Gnedin et al. 2004), namely assuming that the product of the mass in the orbit-averaged radius $\bar{r}$ with radius conserves.
The quantity
\begin{equation}
  \bar{r} = {2 \over T_r} \int_{r_{min}}^{r_{max}} r \, {dr \over v_r},
\end{equation}
is the orbit-averaged radius, and $T_r$ is the radial period.

In the late phase of structure formation, baryons density increase and this produce a sort of coupling among DM and baryons (Klypin et al. 2001; Klypin, Zhao, and Somerville 2002), with a consequent exchange of angular momentum among baryons and DM (see DP12a, Eqs. 11-14).   

The baryon fraction adopted was that obtained by Giodini et al. (2009), $f_B=f_{500}^{stars+gas}=M_{500}^{stars+gas}/M_{500}$\footnote{In this paper the masses, $M_{200}$, and $M_{vir}$, are converted to $M_{500}$ following White (2001), Hu \& Kravtsov (2003), and Lukic et al. (2009).}. 

The density profiles produced by the quoted model are in agreement with those of previous studies (e.g., El-Zant et al. 2001, 2004; 
Nipoti et al. 2004; Jardel \& Sellwood 2009; Mashchenko et al. 2005, 2006, 2007; Romano-Diaz et al. 2008; Governato et al. 2010, 2012; Cole et al. 2011).


\section{Results and discussion}

\begin{center}
\begin{table*}
\caption{Parameters of the clusters. First column: names of the clusters; second: the $M_{200}$ mass given in N13a; third: baryon fraction obtained from Giodini et al. (2009) using the previous $M_{200}$ masses; fourth: slope $\alpha$ obtained by N13b modeling the profile with a gNFW model; fifth: as the previous column but the profile were obtained using our model; sixth: core radius obtained by N13b modeling the profile with a cNFW model; seventh: 
as the previous column but the profile were obtained using our model; eighth: BCG mass and errors obtained from N13a (Table 3, Table 4); ninth: as the previous column but using our model; tenth: value of $j$.} 
{\tiny
\hfill{}
\begin{tabular}{c c c c c c c c c c c c} 
\hline\hline
Cluster & $\log{\frac{M_{200}}{M_{\odot}}}$ & $f_B$ & $(\alpha)_{gNFW}^{(N13b)}$ & $(\alpha)_{gNFW}^{(our)}$ & $(\frac{\log{r_{core}}}{kpc})_{cNFW}^{(N13b)
}$ & $(\frac{\log{r_{core}}}{kpc})_{cNFW}^{(our)}$ & $\frac{M_{BCG}}{10^{11} M_{\odot}}$ & $\frac{M_{BCG}}{10^{11} M_{\odot}}$ &  j  \\ [0.5ex] 
\hline
MS2137 & $14.56^{+0.13}_{-0.11}$ & $0.130^{+0.01}_{-0.008}$ & $0.65^{+0.23}_{-0.30}$ &  $0.66^{+0.11}_{-0.15}$ & $0.45^{+0.38}_{-0.48}$ & $0.44^{+0.19}_{-0.24}$  & 6.56 $\pm 0.59$  & 6.6 $\pm$ 0.29 &$1 j_{*}$\\
A963 & $14.61^{+0.11}_{-0.15}$ & $0.131^{+0.01}_{-0.009}$ & $0.50^{+0.27}_{-0.30}$ &  $0.51^{+0.13}_{-0.15}$ & $0.87^{+0.61}_{-0.71}$ & $0.86^{+0.30}_{-0.35}$ & 10.65 $\pm 0.59$ & 10.7 $\pm$ 0.29 & $1.4 j_{*}$\\
A383 & $14.82^{+0.09}_{-0.08}$ & $0.137^{+0.012}_{-0.01}$ & $0.37^{+0.25}_{-0.23}$ & $0.37^{+0.36}_{-0.32}$ 
 & $0.37^{+0.72}_{-0.64}$ &  $0.37^{+0.12}_{-0.11}$  & 9.18 $\pm 0.59$  & 9.2 $\pm$ 0.29 & $1.75 j_{*}$\\
A611 & $14.92 \pm 0.07$ & $0.140^{+0.012}_{-0.011}$ & $0.79^{+0.14}_{-0.19}$ & $0.79^{+0.07}_{-0.09}$ & $0.47^{+0.39}_{-0.50}$ & $0.47^{+0.19}_{-0.25}$ 
 & 12.25 $\pm 0.59$ &  12.3 $\pm$ 0.29 & $j_{*}/1.5$\\
A2537 & $15.12\pm0.04$ & $0.146^{+0.013}_{-0.013}$ & $0.23^{+0.18}_{-0.16}$ & $0.22^{+0.09}_{-0.08}$ & $1.67^{+0.24}_{-0.23}$ & $1.68^{+0.12}_{-0.11}$
& 13.60 $\pm 0.59$ &  13.5 $\pm$ 0.29 & $2.1 j_{*}$\\
A2667 & $15.16 \pm 0.08$ & $0.147^{+0.016}_{-0.013}$ & $0.42^{+0.23}_{-0.25}$ & $0.41^{+0.11}_{-0.12}$ & $1.29^{+0.49}_{-0.49}$ & $1.30^{+0.24}_{-0.24}$ 
& 7.94 $\pm 0.59$ & 7.8 $\pm$ 0.29 & $1.6 j_{*}$\\
A2390 & $15.34^{+0.06}_{-0.07}$ & $0.153^{+0.017}_{-0.016}$ & $0.82^{+0.13}_{-0.18}$ & $0.82^{+0.06}_{-0.08}$ & $0.30^{+0.53}_{-0.34}$ & $0.30^{+0.26}_{-0.17}$  & 5.26 $\pm 0.59$  & 5.3 $\pm$ 0.29 & $j_{*}/1.4$\\ [1ex]
\hline
\end{tabular}}
\hfill{}
\label{tb:tablename}
\end{table*}
\end{center}

In the introduction, we already pointed out that the slope $\alpha$, both in galaxies and clusters (e.g. Simon et al. 2003, 2005), and clusters (Sa02; Sa04, Sa08; N09, N11, N13) is flatter than simulations and a scatter from object to object is observed. 
In DP12a, and DP12b, we studied the inner slope and its scatter in the case of clusters, and galaxies respectively.
We showed that the role of environment (see also Del Popolo \& Cardone 2012), and the consequent structural differences among structures gives rise to galaxies and clusters with different inner slope. We showed how different baryonic fraction, and angular momentum changes the profile of the structure: structures having larger baryons content (especially in the central region) and larger angular momentum, have flatter inner profiles. Similar to DP12a, we use the model previously described to obtain the mass profile of the seven clusters studied in detail in N13a, N13b. In the two quoted papers, the total (N13a), and the DM density profiles of MS2137, A963, A383, A611, A2537, A1667, and A2390, were determined. Some of the quoted clusters were already studied in previous papers of Newman, and Sand. Namely, MS2137 was previously studied in Sa02, Sa04, and Sa08; A383 was studied in S04, S08, and N11; A963 was studied in S04; A611 was studied in N09. While the analysis for MS2137, and A963 are consistent with S04, and S08, and that of A383 is consistent with N11, the analysis of A611 is different from that of N09. The difference is due to improved stellar kinematics measurements, and a revised redshift of one imaged galaxy.  

In N13a,b, improved data allowed the determination of the stellar mass scale, allowing to produce a more physically consistent 
analysis, reducing the degeneracies among stellar and dark mass, and taking account the BCGs homogeneity. 

The density profiles were calculated by means of the model of DP09, DP12a (summarized in the previous section). 
In order to compare the DM density profiles obtained in our model with the N13b profiles, we need the virial mass of each cluster, the baryon fraction, and the random angular momentum.
The clusters virial mass and the baryon fraction of the clusters studied, 
are given in Table 1 (second and third column). 
The baryonic fraction was obtained following Giodini et al. (2009), after converting the mass $M_{200}$ of the clusters, given in Table 1\footnote{The mass $M_{200}$ of the clusters in Table 1, are taken from N13a.} to $M_{500}$. The shape of the density profile also depends from the angular momentum of the cluster. In the case of clusters of galaxies, differently from the case of galaxies, the role of the ``ordered" angular momentum is not so important, since clusters have very small rotational velocity, while the ``random" angular momentum, $j$, is significant (see DP12a). As in DP09, for the reason now quoted, we assumed the same ``ordered" angular momentum for the clusters studied, fixing it in terms of the spin parameter $\lambda=0.03$, typical value obtained by Gottl\"ober \& Yepes (2007). 
The ``random" angular momentum was fixed by fitting the final DM density profile, similarly to DP12a for the clusters A611, A383, MACS J1423, and RXJ1133, and in DP12b for dwarf galaxies. 

More precisely, 
according to Fig. 1a-c in DP12a, the DM density profile can be written as 
$\rho_{DM}=F(M_{vir}, f_B, j)$. Knowing $M_{vir}$, and $f_B$, we adjusted the value of $j$, so that $\rho_{DM}=F(M_{vir}, f_B, j)$ reproduces the observed clusters profiles. The values of the ``random" angular momentum, expressed in terms of $j_*$
are reported in Table 1 (last column). 
{
$j_*$ is the ``random" angular momentum reproducing the result of DP09 (Fig. 5).
Since the parameter $j_*$ is a fundamental parameter in our model, before going on, we define it better, also showing how the inner slope of the density profile changes with $j_*$.  

As we showed in DP09, the density profile depends on $j$ and on the ordered angular momentum. The value of $j$ which gives rise to the the density profile of a halo of $10^{14} M_{\odot}$ given in DP09 (Fig. 5) defines the typical value $j_*$. 

An approximated value of $j_*$ is given by
\begin{equation}
\frac{j(r)}{(G \rho_c \sigma_0)^{1/2} R_f^2} \simeq \frac{1.5}{\nu^{3/2}} (\frac{r}{R_f})^2
\end{equation}
(Ryden 1988b), where $\sigma_0=\sqrt{\xi(0)}$, being $\xi$ the two point correlation function, the radius, $r$, is proportional to the cluster virial radius ($r \propto M^{1/3}$), $\rho_{\rm c}$ is the critical density, 
$R_{\rm f}$ is the filtering radius, namely the radius at which the power spectrum is filtered\footnote{In our case, on clusters scale.}, and $\nu=\delta/\sigma$ is the peak height, and $\sigma$, the rms density fluctuation (Del Popolo \& Gambera 1996).


In Fig. 1a,b, we show how the density profile changes with $j$ for a fixed value of the baryonic fraction. The solid line in both figures are obtained by using the typical value of ordered angular momentum, which expressed in terms of the spin parameter, is given by $\lambda=0.03$. Differently from spiral galaxies, clusters of galaxies have a negligible rotation support. Ordered angular momentum, and mean rotational velocity are much less than their velocity dispersions ($h<<j$).


The ordered angular momentum used to obtain the solid line in Fig. 1a-1b (obtained by the tidal torque theory as in DP09) is characterized by $\lambda=0.03$\footnote{Note that even if we studies haloes of mass in the range $10^{14}-10^{15} M_{\odot}$, and this imply differences in ordered angular momentum, the spin parameter, $\lambda$ has small differences.}. The baryonic fraction is $F_{B_{\ast}}=M_{b}/M_{500}=0.15$, or $f_d=F_b/f_b \simeq 0.88$ (see McGaugh et al. 2010), and the typical random angular momentum is $j_{\ast}$. The solid line in Fig. 1a-b is obtained using these typical values. 

}

\begin{figure*}
(a)
\subfigure{\includegraphics[width=7.4cm]{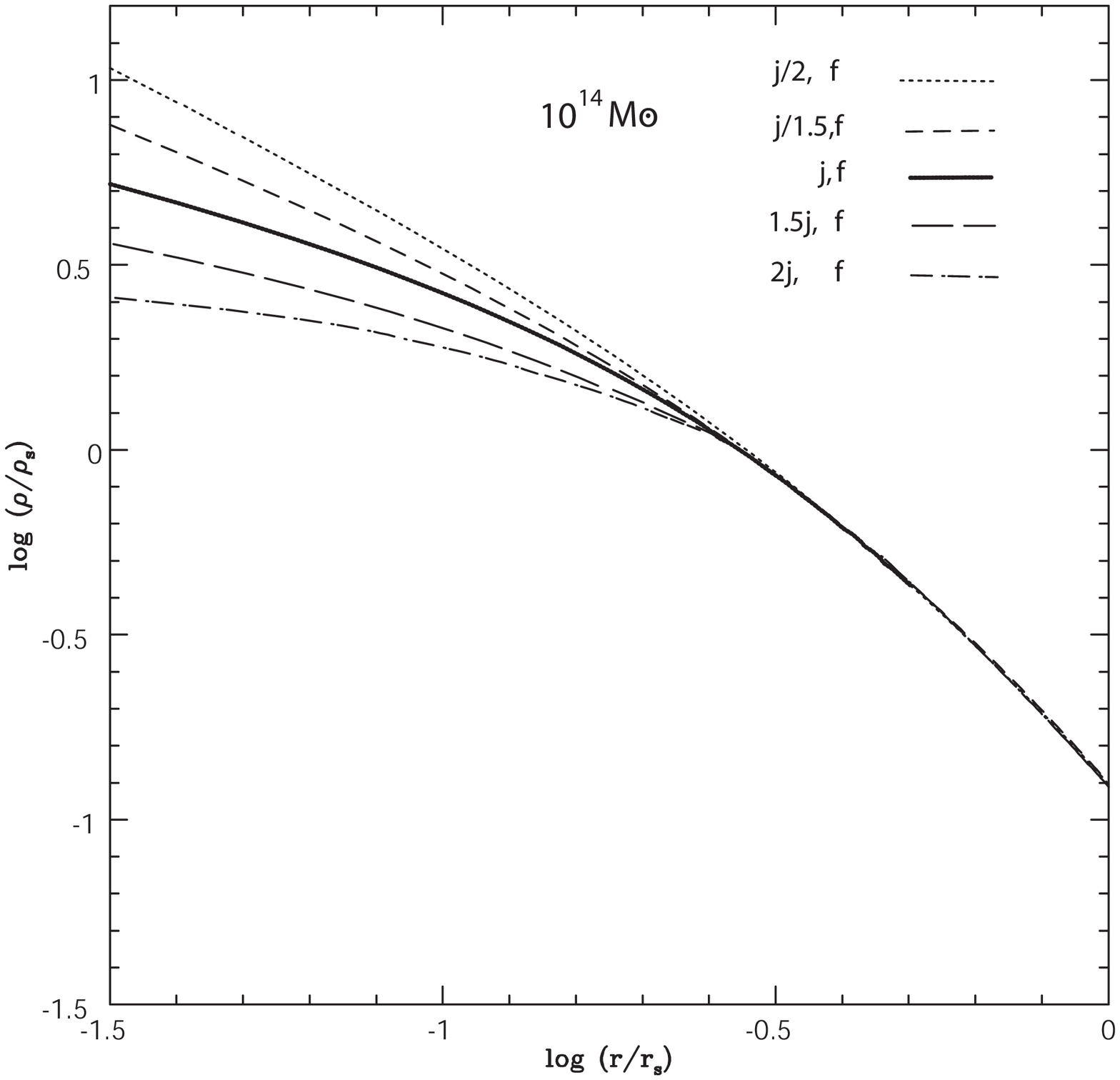}} (b) \goodgap 
\subfigure{\includegraphics[width=7.6cm]{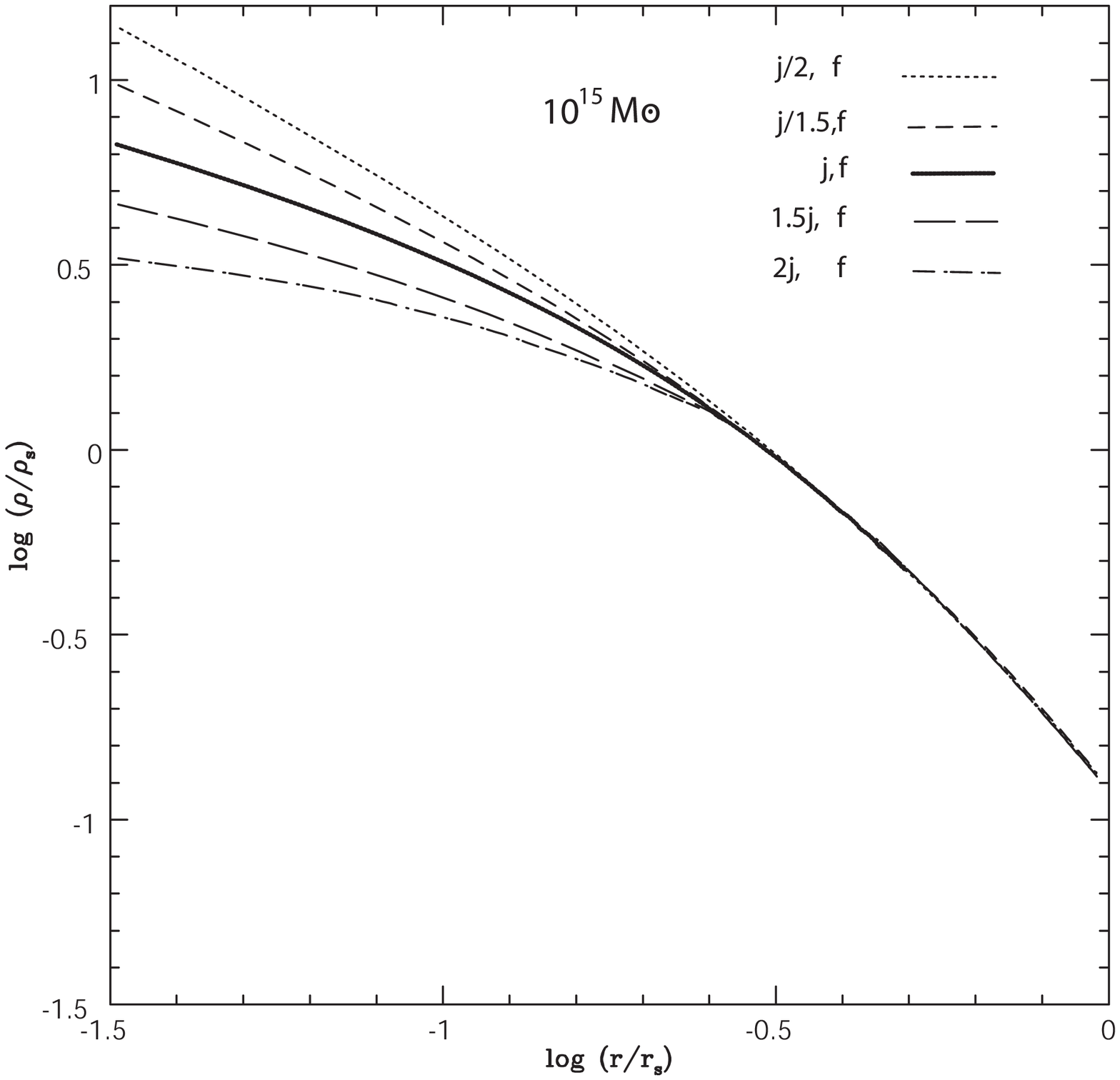}}  
\caption{ Dependence of the density profile with the random angular momentum $j$. Panel (a): change of the density profiles with $j$ in the case of a halo of $10^{14} M_{\odot}$. The typical value of $j$, namely $j=j_{\ast}$, gives rise to the profile given by the solid line, in the case $F_{B_{\ast}}=M_{b}/M_{500}=0.15$ ($f_{d_{\ast}}=0.88$), and $\lambda=0.03$. The upper two lines, namely the dashed and dotted lines, represent the case $j=j_{\ast}/1.5$, and $j=j_{\ast}/2$, respectively. The lower two curves (with respect to the solid line), namely the long-dashed and dot-dashed line, represent the case 
$j=j_{\ast} \times 1.5$, and $j=j_{\ast} \times 2$, respectively. Panel (b): change of the density profiles with $j$ in the case of a halo of $10^{15} M_{\odot}$, similarly to the panel (a).}
\end{figure*}

~\\
~\\

{
Fig. 1 shows how density profile shape changes with changing angular momentum. I want to stress here that even if angular momentum is an important factor to change the shape of the density profile, it is not the primary effect. In our model, angular momentum is not working alone to change the shape of the profiles. As we already reported, DF transfers angular momentum from baryons to DM. Then, the larger is the content of angular momentum of the system, the larger the quantity of the same that can be transferred from baryons to DM, with a consequent further flattening of the profile\footnote{ As shown by several studies, a larger magnitude of angular momentum gives rise to a larger flattening of the profile (Avila-Reese et al. 1998, 2001; Subramanian et al. 2000; Nusser 2001; Hiotelis 2002; Le Delliou \& Henriksen 2003; Ascasibar et al. 2004; Williams et al. 2004; Ascasibar, Hoffman \&
Gottl\"ober 2007).}. 

}

In Fig. 2, we compared the radial density profiles of the DM haloes of the quoted clusters, calculated with the model of this paper with those obtained by N12b. The blue (green) band width represents the DM (total) density profile with 1 $\sigma$ uncertainties calculated by N13b. The dashed bands represent the DM density profiles obtained with the model of the present paper, and the line segment, having slope $\rho \propto r^{-1.13}$\footnote{This is the average slope of the cluster dissipationless CDM simulations in Phoenix project (Gao et al. 2012). }, represents the radius 
$r=0.003-0.03 r_{200}$. The bottom arrows are the three-dimensional half-light radius of the BCG. Since we are mainly interested in the DM distribution, 
we plotted only the density profile of the DM given by our model. However, the total density profile in our model matches that observed by N13b (green  line) equally well as our DM density profile (dashed lines) matches the N13b DM density profile (blue line). Note that in DP09 (Fig. 5), we showed that the total density profile agrees with a NFW profile.

Fig. 2 shows that the density profiles flattens in the inner region of the cluster where the BCG mass starts to be comparable or larger than the DM mass, and this happens in the inner $\simeq 5-10$ kpc (see N13b Fig. 3). At this radii, the total density profile starts to be steeper than the DM density profile, due to the increase of the role of the baryon mass (mainly the BCG mass) at these radii. At radii $\geq 5-10$ kpc the total density profile and the DM profile are very close, since DM is dominating on the baryon component. Outside the inner region of the clusters the slope of the total density profiles (and also DM) are comparable in the different clusters.
At these radii the DM density profile is in agreement with the NFW profile. At radii $\leq 30$ kpc, the total density profile is close to a NFW
profile, and as observed by N13b this, somehow, implies a "tight coordination" among stars distribution and inner DM profile (see the following). 
Since the total mass (composed by the sum of the DM and the baryonic matter) is well described by a NFW profile, in the quoted inner regions of the cluster, and since the baryonic component is dominant in the 5-10 kpc central region, it is logical to expect that the DM density profile is flatter than a NFW one. In fact, as shown by Fig. 2, the inner DM profiles are shallower than a NFW profile ($\alpha <1$), and at the same time exists a scatter in the inner slope between clusters. 

This is better shown in Fig. 3a-b, plotting the slope of the inner profile, $\alpha$, and the core radius, $r_{core}$, 
versus the BCG effective radius $R_e$, which is the radius containing half of the light\footnote{The gNFW is given by
\begin{equation}
\rho_{DM}(r)=\frac{\rho_s}{(r/r_s)^\alpha (1+r/r_s)^{(3-\alpha)}}
\end{equation}
which has a central cusp with $d \log{\rho_{DM}}/d \log{ r \rightarrow 0}$. The cored NFW (cNFW) model is given by
\begin{equation}
\rho_{DM}(r)=\frac{b \rho_s}{(1+b r/r_s) (1+r/r_s)^{2}}
\end{equation}
and has a central density core within $r_s/b$. Both the gNFW, and the cNFW profiles reduce to a NFW profile for $\alpha=1$, and 
$r_{core} \rightarrow 0$.}. 
The slope $\alpha$, and the core radius were obtained by parameterizing the halo as a generalized NFW model (gNFW), and as a cored NFW model (cNFW), respectively. Note that as already stressed by N13b, the choice of the two different parameterizations (gNFW or cNFW) does not affect the result.  
The values of $\alpha$, and $r_{core}$ obtained by N13b are reported in Table 1 (column 4, and 6), and the values of the same parameters obtained in our model are reported also in Table 1 (column 5, 7).

While Fig. 3a is taken from N13b (their Fig. 5)\footnote{In order to have a more readable figure, in our Fig. 3 we chose to reproduce the Fig. 5 of N13b, instead of superimposing the N13b results on ours.}, Fig. 3b plots the same quantities obtained using our model. 

The gray points in the upper part of Fig. 3a, taken from N13b, plots the slopes of clusters similar to those studied in the present paper, obtained in dissipationless N-body simulations (Gao et al. 2012). The dashed horizontal line represents the mean slope. 
The dotted line shows a weak slope change from one cluster to the other.


The colored points with errorbars in fig. 3a-3b, shows that in real clusters, the previous scatter is much larger: clusters having larger BCGs have a shallower inner slope, with respect to clusters having larger BCGs. Similarly, clusters having smaller BGCs have smaller core radii (see the bottom panel).
The dotted lines are least-squares linear fits to the data. 
The previous correlation, namely larger BCG having flatter density DM profiles, is in line with what was previously written. A larger BCG has a larger total mass 
and then the DM mass, $M_{DM}=M_{total}-M_{baryon}$, in the inner regions is less than in a smaller BCG, and consequently the inner DM slope is smaller. The ``constraint" that the total mass has a density profile with $\alpha \simeq 1$, suggests that
the inner profile shape and slope is strictly connected to star formation in the inner part of the cluster. 

A comparison of Fig. 3a with Fig. 3b, shows that slopes obtained with our model are very close to those of the observations.
The result of our model shows the same correlations as the observations, among $\alpha$, $r_{core}$, and $R_e$. The Spearman rank correlation coefficient and the corresponding P-value relative to $\alpha$, and $r_{core}$ obtained from our model, are close to those obtained by N13b. In the case of the $\alpha$-$R_e$ correlation, we have $\rho=-0.58$, $P=0.2$ (N13b obtained $\rho=-0.57$, $P=0.18$), while for the correlation $r_{core}$-$R_e$, we have $\rho=+0.72$, $P=0.07$ (N13b obtained $\rho=+0.71$, $P=0.07$). 


%
%

{ We want to stress that both in N13b, and the present paper, the correlation between the core radius, $r_{core}$, and the BCG effective radius, $R_e$, is strongly dependent on the two data points at largest radii. }

In Fig. 4, we plot $\alpha$ in terms of the BCG mass, $M_{BCG}$\footnote{N13b plotted just the $\alpha$, $r_{core}$-$R_e$ correlation. They just mentioned to its existence.}. 
In the left panel the $\alpha$ values are those obtained in N13b observations, and 
the BCG mass and errors are obtained from N13a (Table 3, Table 4), and are reported in column 8 of Table 1. Columns 9  of Table 1, reports the $M_{BCG}$ obtained by means of our model. Following N13b, the errors in $M_{BCG}$ was assumed equal to 0.07 dex. 
In the right panel, $\alpha$ is plotted in terms of the BCG mass, $M_{BCG}$, obtained using our model (see Table 1). 
The plot shows that clusters with more massive BCGs have flatter density profiles, in agreement with DP12a (Fig. 4b)\footnote{Note that Fig. 4b of DP12a plots the slope dependence on the inner baryonic fraction only in the case $j=j{*}$.}. As already reported, this is expected, since a larger baryonic mass in 5-10 kpc implies a flatter DM density profile, if the total density profile has to be a NFW.
The dotted line is again the least-squares linear fit. In order to estimate the correlation, we calculated the Spearman rank correlation coefficient, which is in this case $\rho=-0.536$, and the P-value $P=0.2$ testifying for a correlation between $\alpha$and $M_{BCG}$.

{As discussed before, and shown in Fig. 1, the DM density profile flattens with increasing $j$. From Fig. 4, we know that the larger is $M_{BCG}$ the flatter is the inner density profile. We then expect another correlation, not plotted, between $M_{BCG}$ and $j$. The larger is $j$, 
the larger is $M_{BCG}$. This because a larger angular momentum gives rise to a flatter DM profile, and since the total mass profile is close to a NFW profile, this implies that the content of baryons, and $M_{BCG}$, is larger.}

The shapes of the density profiles, and the scatter from cluster to cluster shown in Fig. 2, as well as the correlations shown in Fig. 3, 4, can be explained according to our model (DP09, DP12a), as follows. At high $z$, the proto-structure, containing gas and DM, is in its linear phase. In the SIM, the proto-structure is divided into shells, which initially expand with Hubble flow till a maximum radius and then collapse. The DM mass component collapse before the baryonic mass component, and baryons falls in the DM potential wells, radiating part of their energy, and forming clumps which condense into stars (see Li et al. (2010) (Sect. 2.2.2, 2.2.3), De Lucia \& Helmi 2008). In the baryon collapse phase, DM is compressed in the so called ``adiabatic contraction" (AC) (Blumenthal et al. 1986; Gnedin et al. 2004, 2011), and stars form. This dissipational process, happening at $z \geq 2$ (see DP09, Fig. 3, and 5) gives rise to a steep density profile, which constitutes the main structure of the BCG (see also Immeli, Samland, Gerhard, \& Westera 2004; Lackner \& Ostriker 2010), having a scale radius,
$R_e \simeq 30$ kpc, which is similar to the sizes of massive galaxies at high redshift (Trujillo et al. 2006; Williams et al, 2010; Newman et al. 2012). Subsequent merging of satellites with the proto-BCG adds stars to the outer parts of the BCG (e.g., Naab et al. 2009, Laporte et al. 2012).

The clumps formed in the baryons collapse phase, moving to the center are exposed to the dynamical friction (DF) from DM particles. The result is a motion of the DM towards the outer parts of the proto-structure reducing the central DM density (El-Zant 2001, 2004; Nipoti et al. 2004; DP09; Cole et al. 2011).
%
%
Other mechanisms proposed to flatten the DM profile are feedback from AGN (e.g. Martizzi et al. 2012). However, the process seems to be "too effective", since it produces a core of 10 kpc, which is much larger than what observed (Postman et al. 2012). 

In our scenario, one expects a flattening of the DM density profile, and at the same time an anti-correlation of the inner DM slope, $\alpha$, with the 
central baryonic content of the cluster (Nipoti et al. 2004; DP09; DP12a). In more detail, in our model the density profile shape is regulated by angular momentum, baryonic fraction, and virial mass, $\rho_{DM}=F(M_{vir}, f_B, j)$. The effect of angular momentum in shaping the inner profile is larger than that of the baryon fraction, and the role of baryon fraction is larger than that of virial mass. The flattening of the inner slope due to the angular momentum is due to the fact that the shells in a proto-structure having larger angular momentum, tend to remain closer to the maximum radius, and consequently they do not contribute to the central density. In Table 1, the last column, shows for each cluster the angular momentum $j$\footnote{As we already wrote, the ordered angular momentum is equal for all clusters and has a characteristic spin parameter $\lambda=0.03$.}. As an example, A2537, and A2667, have a similar value of virial mass (more correctly $M_{200}$), and baryon fraction, $f_B$. The shallower profile of A2537 is due to its larger 
$j$ (2.1 $j*$ versus $1.6 j_*$ for A2667)\footnote{For sake of precision, we should point out that the low value of the A2537 slope could be produced by the fact that it could be a l.o.s. merger (N13a,b). This produces a shallower profile in lensing analysis. At the same time, A2537 is the largest mass BCG (second-largest when looking at the $R_e$, and according to the previous discussion we expect a shallower profile with respect to the other clusters.}.
The dependence of the slope $\alpha$ on the baryonic mass of the whole cluster, and on the mass contained in the inner 10 kpc was one of the predictions of DP12a (see Fig. 2, and Fig. 4b of the quoted paper).
The flattening of the profile for larger angular momentum was also found by several authors
(Sikivie, Tkachev \& Wang 1997; Avila-Reese et al. 1998; Nusser 2001; Hiotelis 2002; Le Delliou \& Henriksen 2003; Ascasibar et al. 2004; Williams et al. 2004). 
 
The tendency to have flatter profiles for clusters containing larger quantity of baryons, especially at the center, is due to the fact that the larger is the baryonic content of the cluster the larger is the angular momentum transferred from baryons to DM, through DF, with the quoted result that DM particles move away from the cluster center. 

\begin{figure*}
\psfig{file=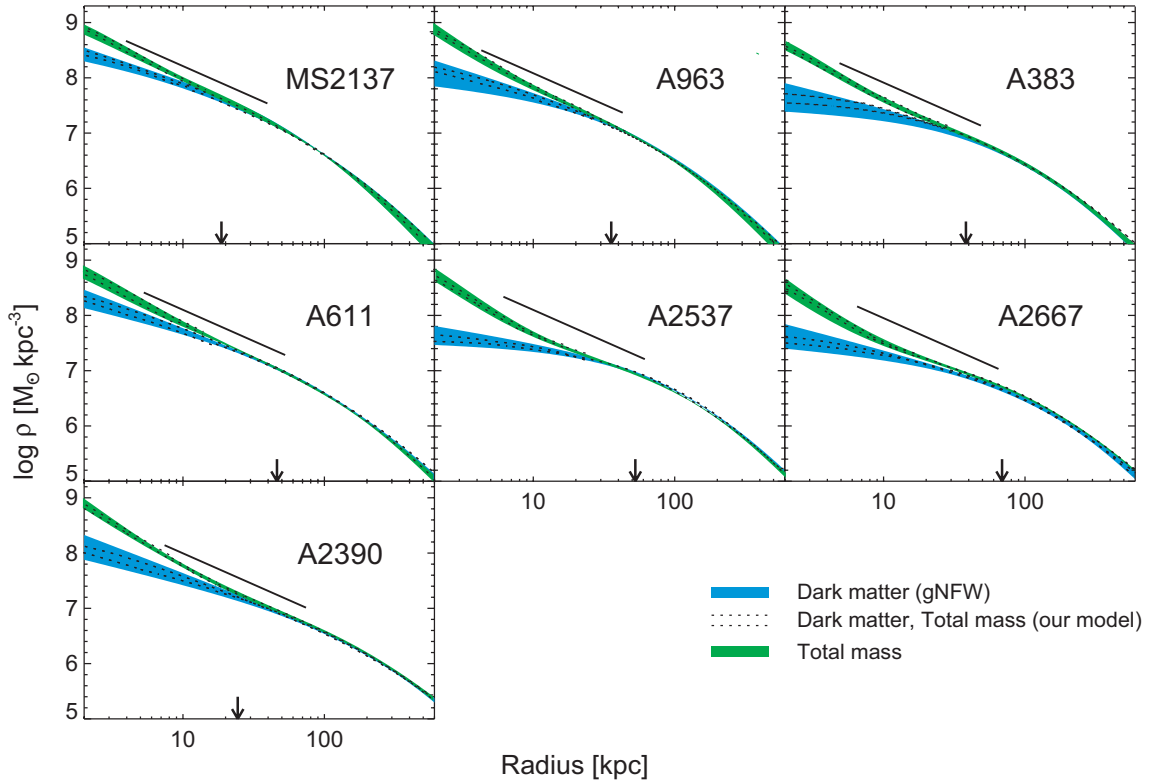,width=15.0cm}
\caption{Density profile of the total mass and DM for the clusters studied in this paper. The bottom blue band (green upper band) represents the DM (total mass) density profile determined by N13b. The band in black dotted lines is the DM density profile obtained in this paper. The band widths represent the 
$1 \sigma$ uncertainty. The bottom arrow, in each panel, is the three-dimensional half-light radius of the BCG. The segment with slope $r^{-1.13}$ spans the radial range $r=0.003-0.03 r_{200}$.}
\end{figure*}

\begin{figure*}
\psfig{file=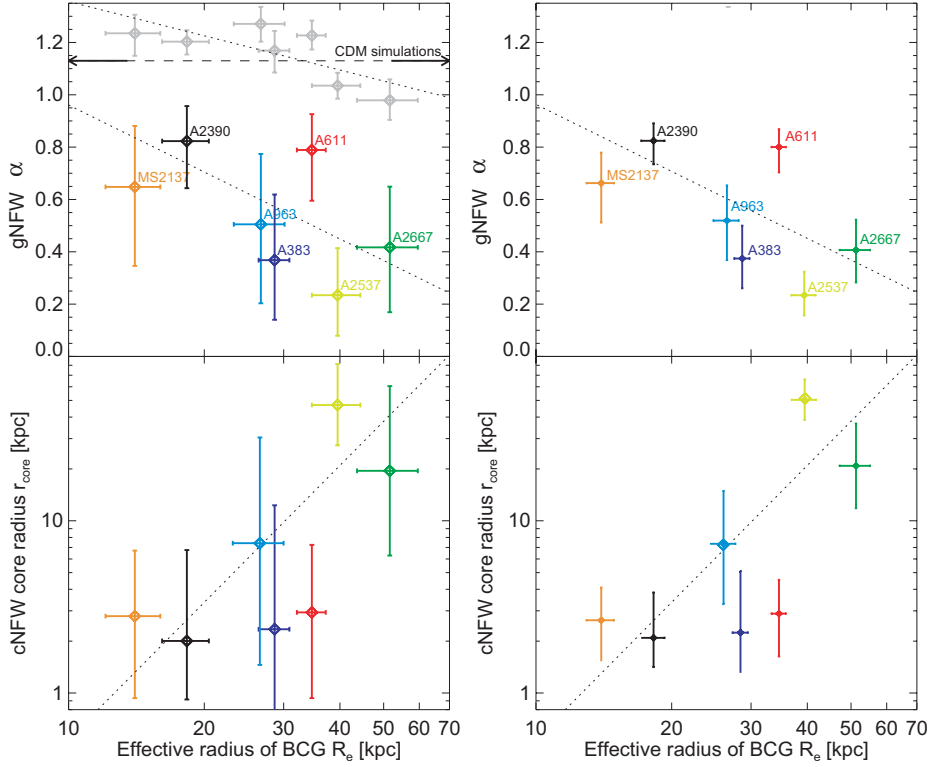,width=13.5cm}
\caption{Correlation among the inner DM profile and the BCG size. Top left panel: the gray points in the upper part of Fig. 3a, taken from N13b, plots the slopes of clusters similar to those studied in the present paper, obtained in dissipationless N-body simulations (Gao et al. 2012). The dashed horizontal line represents the mean slope, and the dotted line shows the weak slope change from one cluster to the other. The colored points are the 
$\alpha$
values vs. $R_e$ for the clusters obtained using the gNFW by N13b. The dotted lines are the least-square fits. Bottom left panel: the core radii, $r_{core}$ of the cNFW model vs. $R_e$. 
Top and bottom right panels: similar to the left panels but giving the results obtained by means of our model.}
\end{figure*}

\begin{figure*}
\psfig{file=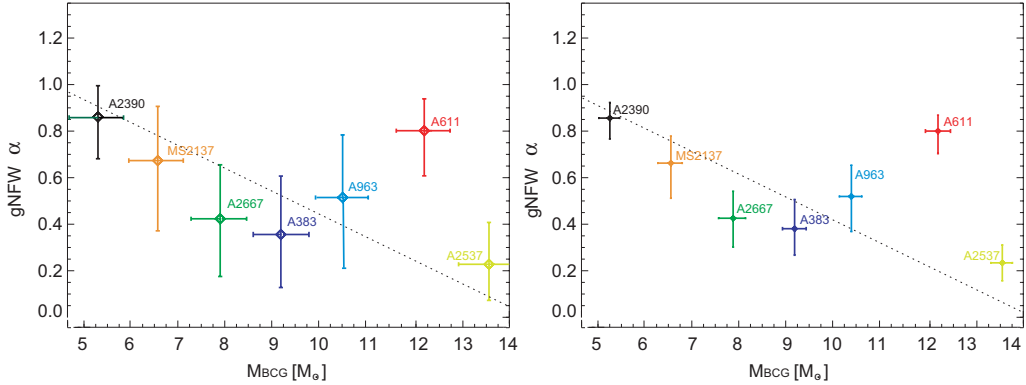,width=13.5cm}
\caption{Correlation among the inner DM profile and the BCG mass. Left panel: the $\alpha$
values vs. $M_{BCG}$ obtained using the gNFW by N13b. The $M_{BCG}$ values were obtained from Table 3, and 4 of N13a. The dotted line is the least-square fit. 
Right panel: similar to the left panel but giving the results obtained by means of our model.}
\end{figure*}

\begin{figure*}
\psfig{file=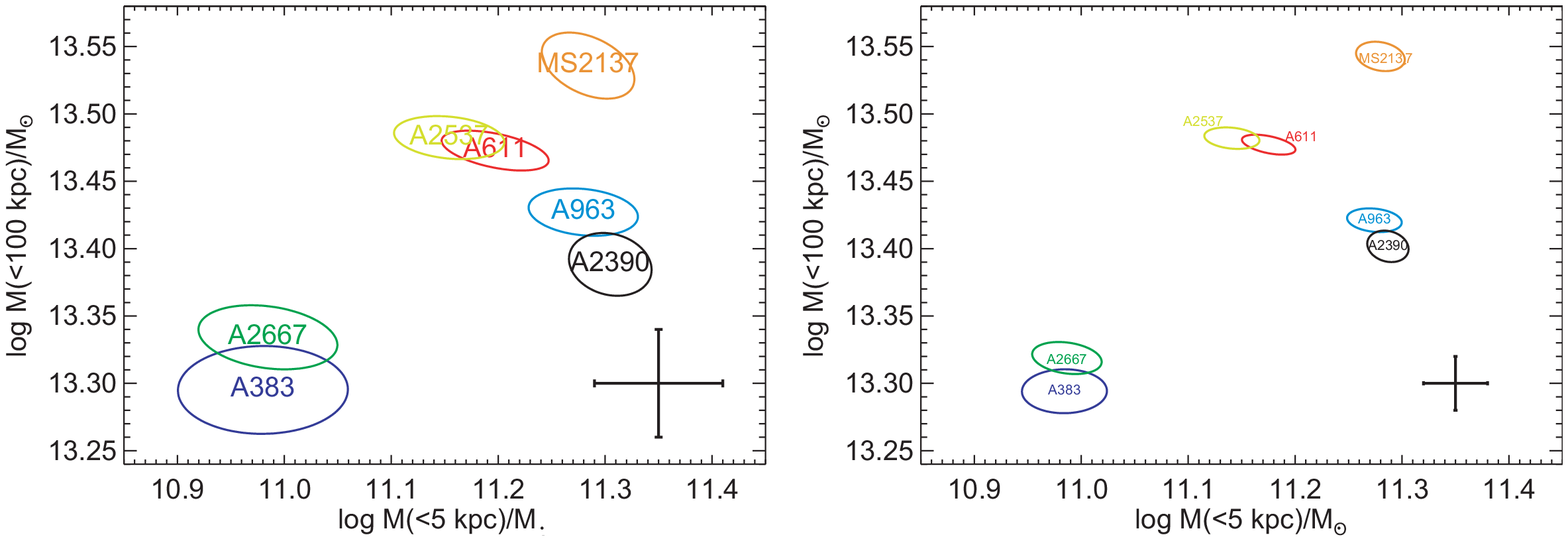,width=13.5cm}
\caption{Mass contained in $r < 100$ kpc vs. the mass contained in 5 kpc. Error ellipses ($\sigma$)  indicated the uncertainty. The left panel refers to N13a data, the right panel to the results of our paper.}
\end{figure*}

The profile steepens with mass because higher density peaks, characterized by larger $\nu$\footnote{$\nu=\delta(0)/\sigma$, where 
$\sigma$ is the mass variance, and $\delta$ the fractional density excess in a shell.} are statistically the forefathers of more massive haloes, and the last have a larger central density contrast. Consequently, a generic shell
will feel a stronger central potential and it will expand less than if the same shell was located in a smaller density peak. The final consequence is a lower quantity of angular momentum acquired in the expansion phase giving rise to haloes more concentrated.


In the case of the clusters studied by N13b no correlation with $M_{vir}$ was found. In our papers DP09, and DP12a, the dependence from the virial mass is slight, and evident only considering large difference in mass. The range of mass of clusters studied in N13b is $\simeq 4 \times 10^{14}- 2 \times 10^{15}M_{\odot}$, and in this small mass range also our model predicts small changes of the values of $\alpha$, in the range $0.62-0.7$ ($\Delta \alpha =0.08$). Moreover, one should take account of the errors in the slope evaluation in N13b, which are larger than the $\Delta \alpha$ in our model. In other terms, taking account of the small mass range considered in N13b,
recalling that in this mass range even our analysis shows a slight correlation with $M_{vir}$, and taking into account the larger errors in their analysis with respect to ours, we may conclude that their result is not in contradiction to ours.


The previous discussion pointed out that in our model the proto-BCG formed at redshift $z \geq 2$, in the dissipative baryonic collapse, and that the further evolution of the BCG was due to later merging of stars on the BCG (e.g., Naab et al. 2009;, Laporte et al. 2012). Subsequently, satellites infalling in the halo produces an ``heating" of DM, and a flattening of the inner slope (El-Zant et al. 2001, 2004; DP09; Cole et al. 2011). 


In this scenario one expects a correlation among the inner mass of the clusters (5 kpc) mainly constituted by stars, and the mass of the core of the cluster (100 kpc), which at the quoted redshift was already formed, and was subsequently subject to little changes (Gao et al. 2004). In Fig. 5, we compare the mass in the central 5 kpc (mainly stars), and that in 100 kpc (mainly DM). Error ellipses ($1 \sigma$) indicates the uncertainty in the observation results (left panel), and in our model (right panel). The left panel is the N13a plot, and the right panel is the result of our model. A correlation with Pearson coefficient $r=0.71$, in our model and 
$r = 0.70$, two-sided $P = 0.08$
in N13a is found. 

The quoted correlation was already discussed in DP13a in the picture of the role of baryons in shaping the DM density profile. 
We showed that the baryonic content was of great importance in shaping the density profile, especially the baryonic content in the central $\simeq 10 $kpc, in the BCG. 
The cluster final configuration, its content in stars, its BCG characteristics, depends from the initial content of baryons, and by formation process. 
These considerations lead to think that the BCG mass and the BCG characteristics should be correlated to the baryonic and cluster mass\footnote{As shown by Whiley et al. (2008) $M_{BCG}\propto M_{cl}^{0.4}$ or $M_{cl}^{0.5}$ according to the feedback model used, and $M_{BCG}\propto M_{cl}^{0.12\pm 0.03}$ for K~band magnitudes inside a diameter of 37~kpc. 
Brough et al. (2008) found $L_{BCG}\propto M_{cl}^{0.11\pm0.10}$ at K~band inside 12$h^{-1}$~kpc (see also Schombert 1988; Edge \& Stewart 1991; Hudson \& Ebeling 1997; Lin \& Mohr 2004; Popesso et al. 2007; Yang et al. 2008; Haarsma et al. 2010).}.

The previous discussion showed that the DM density profile of the clusters studied has $\alpha <1$, and that the total density profile is in agreement with a NFW profile. 

In the past, observations using lensing (Tyson et al. 1998; Smith et al. 2001; Dahle, Hannestad \& Sommer-Larsen 2003; Sa02; Gavazzi et al. 2003; Gavazzi 2005; S04; Brada{\v c} et al. 2008; Limousin et al. 2008), X-ray (Ettori et al. 2002; Arabadjis, Bautz \& Garmire 2002; Lewis, Buote \& Stocke 2003), or combination of strong, weak lensing, and stellar kinematics (e.g. Sa02; Sa04; Sa08; N09, N11) obtained large scatter in the value of $\alpha$. In the case of X-ray observations gave values of $\alpha$ ranging from $\alpha=0.6$ (Ettori et al. 2002) to $\alpha=1.9$ (Arabadjis, Bautz \& Garmire 2002). In the case of lensing, Smith et al. (2001) found $\alpha >1$ for A383, while for the same cluster Sa04, and N11 found $\alpha <1$ combining lensing and stellar kinematics. Tyson, Kochansky \& Dell'Antonio (1998) found $\alpha = 0.57 \pm 0.024$ for Cl 0024+1654, while Kneib et al. (2003), for the same cluster,  found $\alpha \simeq 1$. Sa02, Sa04, and Sa08 found a cored profile for MS2137.3-2353, while Gavazzi et al. (2003, 2005) found different results 
depending from the mass-to-light ratio of the BCG. 

The previous example shows that, in the past, large scatter was obtained in the inner slope of clusters, and in some cases discrepant results were obtained for the same very cluster (see DP12a for a deeper discussion). The quoted scatter and discrepancies, have been attributed to a) degeneracies of $\alpha$ with the concentration parameter, $c$, or, the scale parameter, $r_s$; b) spherical modeling of clusters (Morandi et al. 2010); c) the BCG not taken into account or not properly taken into account; d) difference in the dynamic range in different studies. 
Apart from these reasons, the most trivial reason for discrepancy is that many studies do not specify if the study regards DM or the total density profile, and several times the DM inner slope has been compared to that of the total density profile.  
So, when comparing different results, particular caution must be mind to the radial range considered, and to understand if the paper is studying the \emph{total} density profile or that of the \emph{dark matter} only. 
With this in mind, a comparison of the result of this paper, and those of N13a,b, with more recent observations, usually gives results in agreement. 

In this paper, we are more interested in studying the DM density profiles, so we will make comparisons with previous studies of the DM profile. 
The results concerning the clusters A963, M2137, and A383, are consistent with those of Sa04, Sa08, and N11. 
Lewis et al.(2003), and Zappacosta et al. (2006) studied A2589 and A2029 by using X-rays finding a NFW-like total density $\approx 0.002-0.01 r_{\textrm{vir}}$. Zappacosta et al. (2006) concluded that for any reasonable mass-to-light ratio, the central regions, where the stellar mass is important, are characterized by a shallower DM profile, in agreement with our previous discussions. 
Limousin et al. (2008), and Richard et al. (2009) obtained a value of $\alpha_{\textrm{DM}} = 0.92^{+0.05}_{-0.04}$, considering a fit with a gNFW halo
and BCG stars. This slope is similar to that of A611, and A2390. X-ray studies of a large sample of clusters by Schmidt \& Allen (2007) lead to an estimate of $\langle\alpha_{\textrm{DM}}\rangle = 0.88 \pm 0.29$ (95\% CL). 


Concerning the total density profile, the mean total density profile in our study is $\langle \gamma_{\textrm{tot}} \rangle=1.15 \pm 0.02$\footnote{The average slope of the total mass density profile is calculated as in N13a, as 
$\gamma_{tot}= -\frac{d \log{\rho_{tot}}}{d \log{r}}$. The BCG and the DM halo are distinct components, and $\gamma_{tot}$ is not a directly inferred parameter. 
It is defined by considering a radial interval, $r= 0.003-0.03 r_{200}$, and by fitting a line in the plane of $\log{r}-\log{\rho_{tot}}$.}, similar to that of N13a ($\langle \gamma_{\textrm{tot}} \rangle = 1.16 \pm 0.05 {}^{+0.05}_{-0.07}$) in the radial range $r= r_{200} \times (0.003-0.03)$, in agreement with collisionless DM simulations
(N13a, Section 9)\footnote{It is important to stress that similarly to 
N13a, $\langle \gamma_{\textrm{tot}} \rangle$ is the average slope of the total density profile measured in the radial range $r= r_{200} \times (0.003-0.03)$, and is different from $\alpha_{tot}$, which is the 
asymptotic inner slopes of gNFW models.}.
As shown by N13b, other agreements come when comparing the total density with other observations. Morandi et al. (2011) found $\alpha_{\textrm{tot}} = 0.90 \pm 0.05$ in A1689, for $r \geq 30$ kpc, and Coe et al. (2010) found a that the profile is NFW-like. Postman et al. (2012), Umetsu et al. (2012), and Zitrin et al. (2011) found for MACS~J1206.2-0847 $\alpha_{\textrm{tot}} = 0.96^{+0.31}_{-0.49}$and for A383 $\alpha_{\textrm{tot}} = 1.08 \pm 0.07$, in agreement with our result (NFW-like profile in $r \gtrsim 5-10$~kpc). The stacked density profiles for four clusters studied by Umetsu et al. (2011) gives $\alpha_{\textrm{tot}} = 0.89^{+0.27}_{-0.39}$, when excluding the inner 40~kpc/$h$.

While the total density profile is in agreement with a NFW profile ($\alpha=1$), the inner DM profile is shallower than simulations. Even considering recent simulations (Stadel et al. 2009; Navarro et al. 2010) the minimum slope obtained,  $\alpha \simeq -0.8$ at 120 pc, (Stadel et al. 2009)
is larger than the results of observations, and the scatter in the slope from cluster to cluster is much larger than what found in simulations (see Fig. 2, gray points). If some part of the scatter can be explained, as previously reported, in terms of the limits in techniques (e.g., different dynamic ranges in the studies, the BCG role, simplified modeling of clusters) the difference in slope among some clusters is too big to be explained in this way. 
Since the $\Lambda$CDM model predictions at large scale are in agreement with observations, and since the discrepancies among the $\Lambda$CDM predictions are seen at scales where astrophysical processes are important, probably the discrepancy is due to the lack of baryons (dominant in the inner part of clusters) in dissipationless simulations, as shown in this paper. This astrophysical solution to the quoted discrepancy, is based on the idea that mechanisms "heating" the DM, like "supernova-AGN-driven flattening" (Navarro et al. 1996; Gelato \& Sommer-Larsen 1999; Read \& Gilmore 2005; Mashchenko et al. 2006, 2008; Peirani et al. 2008; Governato et al. 2010; Pontzen \& Governato 2011; Martizzi et al. 2012), or dynamical friction from baryonic clumps (El-Zant  et al. 2001, 2004; Romano-Diaz et al. 2008, 2009; DP09; Cole et al. 2011), are able to reduce the inner density\footnote{Other mechanisms are: interaction of a stellar bar with DM (Weinberg \& Katz 2002; McMillan \& Dehnen 2005);  
decay of binary black hole orbits after galaxies merge (Milosavljevi\'c \& Merritt 2001)}.

In addition to the astrophysical solution to the problem, other more radical solutions, modifying the particles constituting the DM (e.g., Colin, Avila-Reese \& Valenzuela 2000; Sommer-Larsen \& Dolgov 2001; Peebles 2000; Kaplinghat, Knox, \& Turner 2000), modifying the power spectrum at small scales (e.g. Zentner \& Bullock 2003), considering modified gravity (e.g., $f(R)$ theory (Buchdal 1970; Starobinsky 1980); $f(T)$ (see Ferraro 2012), and MOND (Milgrom 1983a,b), have been proposed. 

{
The astrophysical solution, 
however, allows to explain observations in the framework of the $\Lambda$CDM model without the need of modifying the standard cosmological model.
}

\section{Conclusions}

The present paper is an extension and continuation of DP12a in which we used the model introduced in DP09 to understand what is the role of baryonic physics on clusters of galaxies formation. The main goal of the paper was that of studying if the 
DM density profiles of the clusters studied in N13a,b, namely MS2137, A963, A383, A611, A2537, A2667, A2390, and the correlations found by N13a,b, can be reproduced by means of the improved SIM of DP09. To this aim, we calculated, by means of the quoted SIM, the DM and total density profiles of 
the clusters and some correlations observed by N13a,b (also predicted in DP12a), namely: a) the correlation among the inner slope of the DM density profile, $\alpha$, and the effective radius, $R_e$ of the clusters; b) the correlation among the inner slope of the DM density profile, $\alpha$ and the BCG masss, 
$M_{BCG}$; c) the correlation among the core radius $r_{core}$ and $R_e$, and finally d) the correlation among the mass inside 100 kpc, mainly constituted by DM, and the mass inside 5 kpc, mainly constituted by baryons, which indicates that the proto-BCG, and the inner cluster halo were already formed at early times, and lately evolving due to accretion. 

Using the $M_{200}$ mass of the clusters given in N13a, fixing the baryonic fraction of the clusters following Giodini et al. (2009), and fixing the angular momentum as previously discussed, we obtained the DM and total density profiles of the clusters, which are in good agreement with the N13b observations (see our Fig. 2). As already seen in DP12a, the density profiles depend from the baryonic fraction (mainly the central baryonic concentration)
and angular momentum. So, if the baryonic content has an important role in giving rise to the final structure, a fundamental role is also played by 
the dynamics of the clusters constituents. 

The total density profile of the quoted clusters has a mean total density profile $\langle \gamma_{\textrm{tot}}\rangle =1.15 \pm 0.02$,
in the radius range $r=(0.003-0.03) r_{200}$, in agreement with dissipationless simulations and N13a,b, who found 
$\langle \gamma_{\textrm{tot}} \rangle = 1.16 \pm 0.05 {}^{+0.05}_{-0.07}$.

In this range, baryon mass and DM contribute in a significant way to the total mass. 
As seen in Fig. 2, at radii inner than 5-10 kpc, stars dominate the mass distribution, while at outer radii all the mass is fundamentally DM.
This result shows the existence of a "tight coordination" among the inner DM and the stars distribution, as implied by the fact that the NFW-like profile describing the profile is not generated by DM or baryons only but by their mutual action, 
The quoted coordination is further supported
by the correlation among the mass in 5 kpc and that in 100 kpc (Fig. 5), which indicates that the time-scales of formation of the BCG and the inner cluster are similar. 
As discussed in DP12a, the final configuration of a cluster depend from the baryonic content and the formation process. In a hierarchical model of structure formation, we then should expect that the final inner baryonic content and the BCG mass are correlated to the total baryonic and to the mass of the cluster (see Whiley et al. 2008). 

As a further consequence, the inner DM density profile must have a slope $\alpha <1$, since the total mass is the sum of DM and baryons, and since in the inner 5-10 kpc baryons dominate. This is exactly what we see in Fig. 2, and Fig. 3, and Table 1: the slope of all clusters is flatter than $\alpha=1$: the maximum value (including the error) is $\alpha=0.88$ in the case of A2390 and the minimum $\alpha=0.14$ in the case of A2537. 

Moreover, the quoted figures, and Tab. 1, show a large scatter in the inner slope from one cluster to the other, and at the same time (Fig. 3), an anti-correlation among $\alpha$, and $R_e$: clusters hosting larger BCGs have flatter slopes. The anti-correlation is due to the fact that in order total mass have NFW-like profile, clusters having more massive BCGs at their centers must contain less DM in their center. This implies that they must have a flatter DM slope. As discussed in DP12a, the quoted scatter is connected to the environment role. Different baryonic fraction, and angular momentum changes the profile of the structure, namely, larger baryons content, and larger angular momentum give rise to flatter inner slopes. 


The same anti-correlation observed in Fig. 3, is present among $\alpha$ and the BCG mass, $M_{BCG}$. 

The density profiles shape of the clusters studied, and the correlations found by N13a,b, are all well described in our model (DP09, DP12a). The physical picture of the cluster formation is characterized, in agreement with N13a,b conclusions, by a initial dissipative phase giving rise to a steep stellar density profile, followed by a flattening of the DM density profile due to the heating of DM by baryonic clumps collapsing to the cluster center in agreement with (El-Zant et al. 2001, 2004; Romano-Diaz et al. 2008; DP09, DP12ab, Cole et al. 2011). 

The previous picture is also important to solve the Cusp-Core problem, namely the discrepancy among the cuspy inner density profiles of haloes seen in dissipationless simulation of the $\Lambda$CDM model, and the flatter profiles observed. 

Differently from other solutions to the discrepancy, already discussed, the previous solution 
{
allows to explain observations in the framework of the $\Lambda$CDM model 
}
just recognizing that on small scales we are dealing with astrophysics and not cosmology, and that the role of gas at that scales is of not negligible importance.

\section*{Acknowledgements}
We would like to thank the International Institute of Physics in Natal for the facilities and hospitality, and Charles Downing from Exeter University for a critical reading of the paper.

%
%
%

\bibliographystyle{mn2e}
\bibliography{paper}

\begin{thebibliography}{}
\bibitem{} Allen, S. W., Evrard, A. E., \& Mantz, A. B. 2011, ARA\&A, 49, 409
\bibitem{} Antonuccio-Delogu V., Colafrancesco S., 1994, ApJ, 427, 72
\bibitem{} Arabadjis J. S., Bautz M. W., Garmire G. P., 2002, ApJ, 572, 66
\bibitem{} Ascasibar Y., Yepes G., Gottl¨ober S., 2004, MNRAS, 352, 1109
\bibitem{} Astashenok, A. V., and Del Popolo, A., Cosmological measure with volume averaging and the
vacuum energy problem, Class. Quant. Grav. 29 (2012) 085014 [arXiv:1203.2290] [INSPIRE].
\bibitem{} Avila Reese V., Firmani C., Hernandez X., 1998, ApJ, 505, 37
\bibitem{} Avila-Reese, V., Firmani, C., Klypin, A., \& Kravtsov, A. 1999, MNRAS, 310, 527
\bibitem{} Bett P., Eke V., Frenk C. S., Jenkins A., Helly J., Navarro J., 2007, MNRAS, 376, 215
\bibitem{} Blumenthal G. R., Faber S. M., Flores R., Primack J. R., 1986, ApJ, 301, 27
\bibitem{} Bradac M. et al., 2008, ApJ, 681, 187
\bibitem{} Brough, S., Couch, W.~J., Collins, C.~A., Jarrett, T., Burke, D.~J., \& Mann, R.~G. 2008, MNRAS, 385, L103
\bibitem{} Buchdahl, H. A., 1970, MNRAS, 150, 1-8
\bibitem{} Cardone, V. F., \& Sereno, M. 2005, A\&A, 438, 545 2003, ApJ, 593, 26
\bibitem{} Cardone, V. F., Del Popolo, A., Tortora, C., Napolitano, N. R., 2011, MNRAS 416, 1822
\bibitem{} Cardone, V. F., Del Popolo, A., 2012, MNRAS 427, 3176	
\bibitem{} Catelan P., Theuns T., 1996, MNRAS, 282, 436
\bibitem{} Coe, D., Umetsu, K., Zitrin, A., et al. 2012, ApJ, 757, 22
\bibitem{} Cole, D. R., Dehnen, W., \& Wilkinson, M. I. 2011, MNRAS, 416, 1118
\bibitem{} Colin, P., Avila-Reese, V., \& Valenzuela, O. 2000, ApJ, 542, 622
\bibitem{} Dahle H., Hannestad S., Sommer-Larsen J., 2003, ApJ, 588, L73
\bibitem{} de Blok, W. J. G., Bosma, A., \& McGaugh, S. 2003, MNRAS, 340, 657
\bibitem{} de Blok, W. J. G., \& Bosma, A. 2002, A\&A, 385, 816
\bibitem{} De Lucia G., Helmi A., 2008, MNRAS, 391, 14
\bibitem{} de Naray Kuzio R., McGaugh S. S., Mihos J. C., 2009, ApJ, 692, 1321
\bibitem{} de Naray Kuzio R., McGaugh S. S., de Blok W. J. G., 2008, ApJ, 676, 920
\bibitem{} Del Popolo, A., and Gambera, M., Astron. Astrophys. 308 (1996) 373.
\bibitem{} Del Popolo, A., and Gambera, M., Astron. Astrophys. 357 (2000) 809 [astro-ph/9909156] [INSPIRE].
\bibitem{} Del Popolo, A.,  Astron. Rep. 51 (2007) 169 [arXiv:0801.1091]
\bibitem{} Del Popolo, A., 2009, ApJ 698, 2093
\bibitem{} Del Popolo, A.; Kroupa, P., 2009, A\&A 502, 733	
\bibitem{} Del Popolo, A., 2010, MNRAS 408, 1808
\bibitem{} Del Popolo, A., 2011, JCAP 07, 014
\bibitem{} Del Popolo, A., Cardone, V. F.,  2012, MNRAS 423, 1060
\bibitem{} Del Popolo, A., 2012a, MNRAS 424, 38 
\bibitem{} Del Popolo, A., 2012b, MNRAS 419, 971 
\bibitem{} Del Popolo, A., Cardone, V. F., Belvedere, G., 2013, MNRAS 429, 1080
\bibitem{} Del Popolo, A., 2013, AIP Conference Proceedings 1548, 2
\bibitem{} Del Popolo, A.,  Int. J. Mod. Phys. D 23 (2014), 1430005 [arXiv:1305.0456] [INSPIRE].
\bibitem{} Del Popolo, A., MNRAS 336 (2002) 81 [astro-ph/0205449]
\bibitem{} Del Popolo, A., Lima, J. A. S., Fabris, J. C., Rodrigues, D. C., 2014, JCAP 04, 021
\bibitem{} Del Popolo, A., Hiotelis, N., 2014, JCAP 01, 047	
\bibitem{} A. Del Popolo, F. Pace and J.A.S. Lima, Int. J. Mod. Phys. D 22 (2013a) 1350038 [arXiv:1207.5789] 
\bibitem{} A. Del Popolo, F. Pace and J.A.S. Lima, Mon. Not. Roy. Astron. Soc. 430 (2013b) 628 [arXiv:1212.5092] 
\bibitem{} A. Del Popolo, F. Pace, S.P. Maydanyuk, J.A.S. Lima and J.F. Jesus, Phys. Rev. D 87 (2013) 043527 [arXiv:1303.3628]
\bibitem{} Diemand, J., Zemp, M.,Moore, B., Stadel, J., \& Carollo, C.M. 2005, MNRAS, 364, 665
\bibitem{} Edge, A.~C. \& Stewart, G.~C. 1991, MNRAS, 252, 428
\bibitem{} Eisenstein D. J., Loeb A., 1995, ApJ, 439, 250
\bibitem{} El-Zant, A. A., Hoffman, Y., Primack, J., Combes, F.,\& Shlosman, I. 2004, ApJ, 607, L75
\bibitem{} El-Zant, A. A., Shlosman, I., \& Hoffman, Y. 2001, ApJ, 560, 636
\bibitem{} Ettori S., Fabian A. C., Allen S. W., Johnstone R. M., 2002, MNRAS, 331, 635
\bibitem{} Ferraro, R., 2012, AIP Conf. Proc. 1471, 103-110, arXiv:1204.6273v2
\bibitem{} Fillmore J. A., Goldreich P., 1984, ApJ, 281, 1
\bibitem{} Flores, R. A., \& Primack, J. R. 1994, ApJ, 427, L1
\bibitem{} Fukushige, T., Kawai, A., Makino, J., 2004 ApJ 606, 625-634
\bibitem{} Gao, L. Navarro, J. F., Frenk, C. S., et al. 2012, MNRAS 425, 2169
\bibitem{} Gao, L., Loeb, A., Peebles, P. J. E., White, S. D. M., \& Jenkins, A. 2004, ApJ, 614, 17
\bibitem{} Gavazzi R., 2005, A\&A, 443, 793
\bibitem{} Gavazzi R., Fort B., Mellier Y., Pell´o R., Dantel-Fort M., 2003, A\&A, 403, 11
\bibitem{} Gelato, S., \& Sommer-Larsen, J. 1999, MNRAS, 303, 321
\bibitem{} Gentile, G., Salucci, P., Klein, U., Vergani, D., \& Kalberla, P. 2004, MNRAS, 351, 903
\bibitem{} Giodini S. et al., 2009, ApJ, 703, 982
\bibitem{} Gnedin, O. Y., Ceverino, D., Gnedin, N. Y., Klypin, A. A., Kravtsov, A. V., Levine, R., Nagai, D., \& Yepes, G. 2011, arXiv:1108.5736
\bibitem{} Gnedin, O. Y., Kravtsov, A. V., Klypin, A. A., \& Nagai, D. 2004, ApJ, 616, 16
\bibitem{} Gottl\"ober S., Yepes G., 2007, ApJ, 664, 117
\bibitem{} Governato et al. 2010, Nature 463, 203
\bibitem{} Governato, F., Zolotov, A., Pontzen, A., Christensen, C., Oh, S. H., Brooks, A. M., Quinn, T., Shen, S., Wadsley, J., 2012, MNRAS 422, 1231
\bibitem{} Gunn J. E., 1977, ApJ, 218, 592
\bibitem{} Gunn J. E., Gott J. R., 1972, ApJ, 176, 1
\bibitem{} Gustafsson, M., Fairbairn, M., \& Sommer-Larsen, J. 2006, Phys. Rev. D, 74, 123--522
\bibitem{} Haarsma, D.B., et al. 2010, ApJ 713, 1037–-1047
\bibitem{} Hiotelis N., 2002, A\&A, 383, 84
\bibitem{} N. Hiotelis and A. Del Popolo, Astrophys. Space Sci. 301 (2006) 167 [astro-ph/0508531] 
\bibitem{}N. Hiotelis and A. Del Popolo,  Mon. Not. Roy. Astron. Soc. 436 (2013) 163.
\bibitem{} Hoffman Y., Shaham J., 1985, ApJ, 297, 16
\bibitem{} Hoyle F., 1949, in Burger J. M., van der Hulst H. C., eds, IAU and International Union of Theorethical and Applied Mechanics Symposium, Problems of Cosmological Aerodynamics. IAU, Ohio, p. 195
\bibitem{} Hudson, M.~J. \& Ebeling, H. 1997, ApJ, 479, 621
\bibitem{} Immeli, A., Samland, M.,  Gerhard, O.,  \& Westera, P., 2004, A\&A 413, 547
\bibitem{} Jardel, J. R., \& Sellwood, J. A. 2009, ApJ, 691, 1300
\bibitem{} Kaplinghat, M., Knox, L., \& Turner, M. S. 2000, Phys. Rev. Lett., 85, 3335
\bibitem{} Keeton, C. R. 2001, ApJ, 561, 46
\bibitem{} Klypin, A.,  Zhao, H-S., and Somerville R.S., 2002, ApJ 573, 597
\bibitem{} Klypin, A., Kravtsov, A. V., Bullock, J. S., \& Primack, J. R. 2001, ApJ, 554, 903
\bibitem{} Kneib J. P. et al., 2003, ApJ, 598, 804
\bibitem{} Kneib, J.-P., \& Natarajan, P. 2011, A\&ARv, 19, 47
\bibitem{} Lackner, C. N., \& Ostriker, J. P. 2010, ApJ, 712, 88
\bibitem{} Laporte, C. F. P., White, S. D. M., Naab, T., Ruszkowski, M., \& Springel, V. 2012, MNRAS, 424, 747
\bibitem{} Le Delliou M., Henriksen R. N., 2003, A\&A, 408, 27
\bibitem{} Lewis A. D., Buote D. A., Stocke J. T., 2003, ApJ, 586, 135
\bibitem{} Li, Y-S., De Lucia, G., Helmi, A., 2010, MNRAS 401, 2036 [arXiv: 0909.1291] 
\bibitem{} Limousin M. et al., 2008, A\&A, 489, 23
\bibitem{} Lin, Y.T. \& Mohr, J.~J. 2004, ApJ, 617, 879
\bibitem{} Martizzi, D., Teyssier, R., Moore, B., \& Wentz, T. 2012, MNRAS, 422, 3081
\bibitem{} Mashchenko, S., Couchman, H. M. P., \& Wadsley, J. 2006, Nature, 442, 539
\bibitem{} Mashchenko, S., Wadsley, J., \& Couchman, H. M. P. 2008, Science, 319, 174
\bibitem{} Mashchenko, S., \& Sills, 2005, ApJ, 619, 258
\bibitem{} McMillan P. J., Dehnen W., 2005, MNRAS, 363, 1205
\bibitem{} Milgrom, M. 1983b, ApJ 270, 371-389
\bibitem{} Milgrom, M., 1983a, ApJ 270, 365-370
\bibitem{} Milosavljevi´c M., Merritt D., 2001, ApJ, 563, 34
\bibitem{} Mo, H. J., Mao, S., \& White, S. D. M. 1998, MNRAS, 295, 319
\bibitem{} Moore, B. 1994, Nature, 370, 629
\bibitem{} Moore, B., Governato, F., Quinn, T., Stadel, J., \& Lake, G. 1998, ApJ, 499, L5
\bibitem{} Morandi A., Pedersen K., Limousin M., 2010, ApJ, 713, 491
\bibitem{} Naab, T., Johansson, P. H., \& Ostriker, J. P. 2009, ApJ, 699, L178
\bibitem{} Navarro et al. 2010, MNRAS 402, 21–34
\bibitem{} Navarro, J. F., Frenk, C. S., \& White, S. D. M. 1996, ApJ, 462, 563
\bibitem{} Navarro, J. F., Frenk, C. S., \& White, S. D. M. 1997, ApJ, 490, 493 
\bibitem{} Navarro, J. F., Ludlow, A., Springel, V., et al. 2010, MNRAS, 402, 21
\bibitem{} Navarro, J. F., et al. 2004, MNRAS, 349, 1039
\bibitem{} Newman, A. B., Ellis, R. S., Bundy, K., \& Treu, T. 2012, ApJ, 746, 162
\bibitem{} Newman, A. B., Treu, T., Ellis, R. S., Richard, J., Sand, D. J., 2013, ApJ 765, 25
\bibitem{} Newman, A. B., Treu, T., Ellis, R. S., Sand, D. J., Nipoti, C., Richard, J., Jullo, E., 2013, ApJ 765, 24
\bibitem{} Newman, Andrew B.; Treu, Tommaso; Ellis, Richard S.; Sand, David J., 2011, ApJ 728, 39
\bibitem{} Newman, Andrew B.; Treu, Tommaso; Ellis, Richard S.; Sand, David J.; Richard, Johan; Marshall, Philip J.; Capak, Peter; Miyazaki, Satoshi, 2009, ApJ 706, 1078
\bibitem{} Nipoti, C., Treu, T., Ciotti, L., \& Stiavelli, M. 2004, MNRAS, 355, 1119
\bibitem{} Nusser A., 2001, MNRAS, 325, 1397
\bibitem{} Oh S-H., Brook C., Governato F., Brinks E., Mayer L., de Blok W. J. G., Brooks A., Walter F., 2010, AJ 142, 24 
\bibitem{} Padmanabhan T., 1993, Structure formation in the Universe. Cambridge Univ. Press, Cambridge
\bibitem{} Peebles P. J. E., 1969, ApJ, 155, 393
\bibitem{} Peebles, P. J. E. 2000, ApJ, 534, L127
\bibitem{} Peirani, S., Kay, S., \& Silk, J. 2008, A\&A, 479, 123
\bibitem{} Pontzen, A., \& Governato, F. 2012, MNRAS, 421, 3464
\bibitem{} Popesso, P., Biviano, A., B{\"o}hringer, H., \& Romaniello, M. 2007, A\&A, 464, 451
\bibitem{} Postman, M., Lauer, T. R., Donahue, M., et al. 2012, ApJ, 756, 159
\bibitem{} Postman, M., et al. 2012, arXiv:1205.3839
\bibitem{} Power, C., Navarro, J. F., Jenkins, A., Frenk, C. S., White, S. D. M., Springel, V., Stadel, J., \& Quinn, T. 2003, MNRAS, 338, 14
\bibitem{} Read, J. I., \& Gilmore, G. 2005, MNRAS, 356, 107
\bibitem{} Richard, J., Pei, L., Limousin, M., Jullo, E., \& Kneib, J. P. 2009, A\&A, 498, 37
\bibitem{} Ricotti, M., 2004, MNRAS 2003, 344, 1237
\bibitem{} Ricotti, M., Pontzen, A.,  and Viel, M.,  2007, ApJ 663, 53
\bibitem{} Ricotti, M., and Wilkinson, M.I., 2004, MNRAS 353, 867 
\bibitem{} Romano-Diaz, E., Shlosman, I., Heller, C.,\& Hoffman, Y. 2009, ApJ, 702, 1250
\bibitem{} Romano-Diaz, E., Shlosman, I., Hoffman, Y.,\& Heller, C. 2008, ApJ, 685, L105
\bibitem{} Ryden B. S., 1988, ApJ, 329, 589
\bibitem{} Ryden, B. S., \& Gunn, J. E. 1987, ApJ, 318, 15
\bibitem{} Sand D.~J., Treu T., Ellis R.~S., 2002, ApJ, 574, L129 
\bibitem{} Sand D.~J., Treu T., Smith G.~P., Ellis R.~S., 2004, ApJ, 604, 88 
\bibitem{} Sand, D. J., Treu, T., Ellis, R. S., Smith, G. P., Kneib, J.-P, 2008, ApJ 674, 711
\bibitem{} Schaefer B. M., 2009, Int. J. Mod. Phys. D, 18, 173
\bibitem{} Schmidt R. W., Allen S. W., 2007, MNRAS, 379, 209
\bibitem{} Schombert, J.~M. 1988, ApJ, 328, 475
\bibitem{} Sharma S., Steinmetz M., 2005, ApJ, 628, 21
\bibitem{} Sikivie, P., Tkachev, I. I, \& Wang, Y. 1997, Phys. Rev. D, 56, 1863
\bibitem{} Simon J. D., Bolatto A. D., Leroy A., Blitz L., 2003, ApJ, 596, 957 
\bibitem{} Simon J. D., Bolatto A. D., Leroy A., Blitz L., Gates E. L., 2005, ApJ, 621, 757
\bibitem{} Smith G. P., Kneib J., Ebeling H., Czoske O., Smail I., 2001, ApJ, 552, 493
\bibitem{} Sommer-Larsen, J., \& Dolgov, A. 2001, ApJ, 551, 608
\bibitem{} Span\'o M., Marcelin M., Amram P., Carignan C., Epinat B., Hernandez O., 2008, MNRAS, 383, 297
\bibitem{} Stadel J., Potter D., Moore B., Diemand J., Madau P., Zemp M., Kuhlen M., Quilis V., 2009, MNRAS, 398, 21
\bibitem{} Starobinsky, A. A. (1980). Physics Letters B 91: 99-102
\bibitem{} Treu, T., \& Koopmans, L. V. E. 2002, ApJ, 575, 87
\bibitem{} Trujillo, I., et al. 2006, ApJ, 650, 18
\bibitem{} Tyson J. A., Kochanski G. P., dell'Antonio I. P., 1998, ApJ, 498, L107
\bibitem{} Umetsu, K., Broadhurst, T., Zitrin, A., et al. 2011, ApJ, 738, 41
\bibitem{} Weinberg M. D., Katz N., 2002, ApJ, 580, 627
\bibitem{} Whiley, I.~M., Aragon-Salamanca, A., {De Lucia}, G., {et~al.} 2008, MNRAS, 387, 1253
\bibitem{} White S. D. M., 1984, ApJ, 286, 38
\bibitem{} Williams L. L. R., Babul A., Dalcanton J. J., 2004, ApJ, 604, 18
\bibitem{} Williams, R. J., Quadri, R. F., Franx, M., van Dokkum, P., Toft, S., Kriek, M., \& Labb\'e, I. 2010, ApJ, 713, 738
\bibitem{} Yang, X., Mo, H.~J., \& {van den Bosch}, F.~C. 2008, ApJ, 676, 248
\bibitem{} Zappacosta L., Buote D. A., Gastaldello F., Humphrey P. J., Bullock J., Brighenti F., Mathews W., 2006, ApJ, 650, 777
\bibitem{} Zentner, A. R., \& Bullock, J. S. 2003, ApJ, 598, 49
\bibitem{} Zitrin, A., Broadhurst, T., Coe, D., et al. 2011, ApJ, 742, 117
\end{thebibliography}

\label{lastpage}

{}
\end{document}